\documentclass[prd,aps,floatfix,nofootinbib,twocolumn,tightenlines]{revtex4}
\usepackage{dcolumn} 
\usepackage{amsmath}
\usepackage{latexsym}
\usepackage{graphicx}
\usepackage{bm}
\usepackage{color}

\def\ttl#1{{``#1''}}
\def\ttl#1{}

\def\dlbl#1{\marginpar{\vskip -5ex {\small \underline{#1}} } \hskip -.75ex}
\def\ulbl#1{\marginpar{\vskip 2ex {\small \underline{#1}} } \hskip -.75ex}
\def\dlbl#1{}
\def\ulbl#1{}

\def\floatcaption#1#2{ \caption{ #2 \ [#1] \label{#1}} }
\def\floatcaption#1#2{ \caption{#2 \label{#1}} }

\def\dttl#1#2{\multicolumn{1}{c}{\hspace{#2ex} #1}}

\newcommand{\bee}{\begin{equation}}
\newcommand{\ee}{\end{equation}}
\newcommand{\beea}{\begin{eqnarray}}
\newcommand{\eea}{\end{eqnarray}}
\def\NON{\nonumber\\}

\def\a{\alpha}
\def\b{\beta}

\def\d{\delta}
\def\f{\phi}                    
\def\g{\gamma}

\def\k{\kappa}
\def\l{\lambda}
\def\m{\mu}
\def\n{\nu}
\def\o{\omega}
\def\p{\pi}                     
\def\r{\rho}                    
\def\t{\tau}

\def\x{\xi}
\def\z{\zeta}
\def\D{\Delta}

\def\O{\Omega}

\def\cp{{\cal P}}

\def\tr{{\rm tr}\,}

\def\Re{{\rm Re\,}}

\long \def \blockcomment #1\endcomment{}
\def\det{{\rm det}}

\def\tQ{\tilde{Q}}
\def\tO{\tilde{\O}}
\def\tV{\tilde{V}}
\def\bQ{\overline{Q}}
\def\bO{\overline{\O}}
\def\bV{\overline{V}}
\def\bQ{\bar{Q}}
\def\bO{\bar{\O}}
\def\bV{\bar{V}}
\def\hm{\hat{\mu}}  
\def\hn{\hat{\nu}}

\begin{document}

\title{Suppressing dislocations in normalized hypercubic smearing}
\author{Thomas DeGrand}
\affiliation{Department of Physics,
University of Colorado, Boulder, CO 80309, USA}
\author{Yigal Shamir}
\author{Benjamin Svetitsky}
\affiliation{Raymond and Beverly Sackler School of Physics and Astronomy,
  Tel~Aviv University, 69978 Tel~Aviv, Israel\\ }

\begin{abstract}
Normalized hypercubic smearing improves the behavior of dynamical
Wilson-clover fermions, but has the unwanted side effect that
it can occasionally produce spikes in the fermion force.
These spikes originate in the chain rule connecting the derivative
with respect to the smeared links to the derivative with respect to
the dynamical links, and are associated with the presence
of dislocations in the dynamical gauge field.
We propose and study an action designed to suppress these dislocations.
We present evidence for improved performance of
the hybrid Monte Carlo algorithm.
A side benefit is improvement in the properties of valence chiral fermions.
\end{abstract}

\maketitle

\section{Introduction \label{sec:intro}}
\dlbl{sec:intro}

Smeared links are widespread in present-day lattice gauge theory
simulations.  A smeared link $V_{x,\m}\in {\rm U}(N_c)$
is a parallel transporter from $x$ to $x+\hm$
that is constructed from the dynamical gauge field $U_{x,\m}\in {\rm SU}(N_c)$
in the vicinity of the lattice sites $x$ and $x+\hm$.  The replacement
of the dynamical, or thin, links $U_{x,\m}$ in the fermion action
by smeared, or fat, links $V_{x,\m}$ typically leads to a reduction
of the discretization error.  Intuitively, the fat links $V_{x,\m}$
provide a smoother background for the fermions to propagate in,
resulting in a more continuum-like behavior and thus smaller lattice
artifacts.

In our own work on lattice gauge theory with fermions in higher
representations we have been using normalized hypercubic (nHYP)
smeared links in the Wilson-clover fermion action
\cite{Hasenfratz:2001hp,Hasenfratz:2007rf}.
We have indeed observed
much reduced discretization effects.  This allowed us to
reach deeper into strong coupling, as well as to keep the
clover term at its tree-level value $c_{SW}=1$
\cite{DeGrand:2010na,Shamir:2010cq}.

Our simulations with dynamical fermions were carried out with the standard
Hybrid Monte Carlo (HMC) algorithm \cite{HMC}.
We use the familiar tools to accelerate the
molecular dynamics (MD) integration: an additional heavy pseudofermion field
as suggested by Hasenbusch~\cite{Hasenbusch:2001ne}; multiple time
scales for nested MD integration levels~\cite{Urbach:2005ji};
and a second-order Omelyan integrator~\cite{Takaishi:2005tz}.

This arsenal of techniques lends itself to many variations.
As an example, one can try to improve the overall performance
of the algorithm by using several intermediate Hasenbusch masses,
each with its own pseudofermion action,
at the same MD integration level.  The idea is that, unlike
the number of nested integration levels, which can only be changed
discretely, the masses of the additional heavy pseudofermions
can be tuned continuously, allowing for a more efficient optimization.
This approach turned out to be
successful for domain-wall fermions \cite{Yin:2011np,Arthur:2012opa}.

We have experimented with these improvement schemes, but in many cases we
have been stymied by continued low acceptance.
Examination of our results suggests that the explanation of the problem lies in our smearing procedure.
In the domain-wall simulations of the RBC and UKQCD collaborations
\cite{Yin:2011np,Arthur:2012opa}, smeared links have not been used, and
the fermion force was obtained by directly differentiating
the pseudofermion action with respect to the dynamical links $U_{x,\m}$.
In our simulations, on the other hand,
we first differentiate the pseudofermion
action with respect to the nHYP links $V_{x,\m}$, and then apply
the chain rule that is needed to convert the derivative with respect
to the nHYP links into a derivative with respect to the dynamical (thin)
links $U_{x,\m}$.
As will be clear below, this chain rule is the origin of our difficulties.

In this paper, we describe the problem and propose and study a remedy.
Reduced acceptance is a result of spikes in the force;
these appear in the chain-rule calculation because of dislocations
in the dynamical gauge field, which produce large derivatives
through the normalization step in the smearing.
Our remedy is a new term in the gauge action, which suppresses
these dislocations.
This tames the fluctuations of the fermion force.

In Sec.~\ref{sec:chain} we present the evidence for the connection between
the fat-to-thin chain rule and the acceptance of the HMC algorithm.
In Sec.~\ref{sec:nds} we present the new term in the gauge
action, designed to suppress dislocations.
In Sec.~\ref{sec:MD} we display the resulting improvement in performance
of the HMC algorithm.  We have also observed a side benefit---better behavior
of chiral valence fermions, as we report in Sec.~\ref{sec:ov}.
We conclude with a discussion in Sec.~\ref{sec:conc}.

Of the smearing techniques in widespread use,
Highly Improved Staggered Quarks, known as HISQ \cite{HISQ,HISQ2},
also make use of a normalization (i.e., reunitarization) step
[see Eq.~(\ref{reu}) below].
Our technique can be applied to them as well.
Stout links \cite{Morningstar:2003gk} do not: The smeared link is
an analytic function of the thin links,
and the problem we encounter with nHYP smearing is avoided.
In favor of nHYP links, we recall their advantage,
that the smearing range is small.  Stout smearing, on the other hand,
is usually repeated several times, giving rise to
a less local fermion action.
Local hypercubic geometry can be combined with the analytic stout recipe
as HEX \cite{HEX} or sHYP \cite{Hasenfratz:2007rf} smearing.

\begin{table}[t]
\floatcaption{tab:configs}{Ensembles used in this study
and overall performance.
The parameter $\g$ is the coefficient of the new term in the gauge action,
see Eqs.~(\ref{dsa}) and~(\ref{gamma}) below.
$\b$ is the usual plaquette coupling, and $\k$ the hopping parameter.
$n_1$, $n_2$ and $n_g$ are the number of steps
of the three MD integration levels.
The last column gives the acceptance rate.
In all cases the lattice size is $L^3\times T = 12^3\times 24$,
and the trajectory length is set to unity.
The number of configurations in each ensemble ranges between 400 and 800.}
\begin{center}
\begin{ruledtabular}
\begin{tabular}{lll lll l}
$\g$ & $\b$ & $\k$ & $n_1$ & $n_2$ & $n_g$ & acc. \\
\hline
0     & 9.6  & 0.1292 & 16 & 2 & 6 & 86\% \\
0     & 9.65 & 0.129  & 16 & 2 & 6 & 92\% \\
0.125 & 7.8  & 0.130  & 12 & 1 & 5 & 82\% \\
0.25  & 5.6  & 0.130  & 12 & 1 & 5 & 92\% \\
\end{tabular}
\end{ruledtabular}
\end{center}
\end{table}

\begin{table*}[t]
\floatcaption{tab:spec}{Physical properties of the ensembles.
$m_q$ is the quark mass as determined from the unimproved axial
Ward identity \cite{DeGrand:2010na}.
$r_1$ is the larger Sommer scale \cite{Sommer:1993ce,Bernard:2000gd},
while $m_\p$ and $m_\r$ are the pseudoscalar and vector meson masses
respectively. All results are given in lattice units.}
\begin{center}
\begin{ruledtabular}
\begin{tabular}{lldddd}
  $\g$ & $\b$ &
  \dttl{$m_q$}{5}  &
  \dttl{$r_1$}{3}  &
  \dttl{$r_1 m_\p^2/m_q$}{1}  &
  \dttl{$(m_\p/m_\r)^2$}{3}
\\ \hline
0     & 9.6  & 0.0588(4) & 3.04(6)  &  9.1(2) & 0.43(1) \\
0     & 9.65 & 0.0471(4) & 3.53(14) & 12.2(5) & 0.44(1) \\
0.125 & 7.8  & 0.0484(5) & 3.12(5)  &  9.7(2) & 0.40(1) \\
0.25  & 5.6  & 0.0575(3) & 3.22(5)  & 10.7(2) & 0.48(1) \\
\end{tabular}
\end{ruledtabular}
\end{center}
\end{table*}

\section{Chain rule and acceptance rate \label{sec:chain}}
\dlbl{sec:chain}

Schematically, an nHYP link $V$ is constructed as
\ulbl{reu}
\begin{equation}
  V = \cp(\O) \equiv \O Q^{-1/2} \ ,
\label{reu}
\end{equation}
where\footnote{%
  In the numerical implementation we modify $Q = \O^\dagger \O + \z$,
  with $\z=10^{-6}$, to avoid accidental crashes.
}
\ulbl{Q}
\begin{equation}
  Q = \O^\dagger \O \ ,
\label{Q}
\end{equation}
and $\O$ is a weighted sum over paths (the precise definition will be
given in the next section).  Let $S$ be a pseudofermion action
that depends explicitly only on the fat links, and suppose that $U$ is
one of the thin links on which the weighted sum $\O$ depends.
The thin-link force $\partial S/\partial U$ is related to
the fat-link force $\partial S/\partial V$ via the ``fat-to-thin'' chain rule,
\ulbl{fatforce}
\begin{subequations}
\label{fatforce}
\begin{eqnarray}
  \frac{\partial S}{\partial U}
  &=& \sum_{i,j=1}^{N_c} \left(
  \frac{\partial S}{\partial V_{ij}} \frac{\partial V_{ij}}{\partial U}
  +\frac{\partial S}{\partial V_{ij}^*} \frac{\partial V_{ij}^*}{\partial U}
  \right)\ ,
\label{fatforcea}\\
  \frac{\partial V}{\partial U}
  &=& \frac{\partial \O}{\partial U}\, Q^{-1/2}
      + \O\, \frac{\partial Q^{-1/2} }{\partial U} \ .
\label{fatforceb}
\end{eqnarray}
\end{subequations}
Clearly, even if the fat-link force is well-behaved, small eigenvalues of $Q$
can generally lead to large terms in the thin-link force.
For $Q$ to have an exceptionally small eigenvalue, the dynamical
gauge field needs to be rough, or, loosely speaking,
a dislocation should be present.
At the same time, one should keep in mind that a locally rough gauge field
does not always give rise to such exceptionally small eigenvalues.
We will return to these considerations in more detail below.

The model we used for our tests is an SU(4) gauge theory with
two Dirac fermions in the two-index antisymmetric (sextet) representation.
The gauge action is the usual Wilson plaquette action
\ulbl{Splaq}
\begin{equation}
  S_{\textrm{plaq}} = \frac{\b}{2N_c}\; \Re \tr \sum_x \sum_{\m\ne\n}
  (1 - U_{x,\m} U_{x+\hm,\n} U^\dagger_{x+\hn,\m} U^\dagger_{x,\n}) \ ,
\label{Splaq}
\end{equation}
with $N_c=4$, to which we add a new piece, to be described below,
with coefficient $\g$.
As already mentioned we are using the Wilson-clover fermion action
with nHYP links, with hopping parameter $\k$, and with $c_{SW}=1$.
We have one Hasenbusch
mass $\m=0.2$ that effectively separates high- and low-momentum modes in the
fermion determinant.
The total pseudofermion action is thus
\ulbl{Spf,Slow,Shigh}
\begin{eqnarray}
  S_{\textrm{pf}} &=& S_{\textrm{low}} + S_{\textrm{high}} + S_{\textrm{eo}} \ ,
\label{Spf}\\
   S_{\textrm{low}} &=& \f^\dagger_1\, \frac{1}{M}\, (MM^\dagger + \m^2) \,
   \frac{1}{M^\dagger} \,\f_1 \ ,
\label{Slow}\\
   S_{\textrm{high}} &=& \f^\dagger_2\, \frac{1}{M^\dagger M + \m^2} \,\f_2 \ ,
\label{high}
\end{eqnarray}
where $\phi_1$ and $\phi_2$ are two independent pseudofermion fields.
$S_{\textrm{eo}}$ is the additional pure-gauge term resulting
from even-odd preconditioning, while $M$ is the even-odd preconditioned fermion matrix.
The force due to $S_{\textrm{low}}$, which is sensitive to the small eigenvalues of $M$,
is integrated in the outer MD level of the Omelyan integrator
with $n_1$ steps per trajectory.  The next level, with $n_2$ steps,
integrates the force due to $S_{\textrm{high}}$,
which is sensitive to the large eigenvalues of $M$,
as well as the force coming from $S_{\textrm{LU}}$.
All these terms depend on the nHYP links.
The force due to the new term, which we will introduce in the next section,
is also integrated at this level.
Finally, the force due to the Wilson plaquette action~(\ref{Splaq})
is integrated in the innermost level with $n_g$ steps.

\begin{table*}[t]
\floatcaption{tab:impl}{Maximal and average impulse per trajectory before
and after the chain rule.  Results are shown for accepted and for rejected
trajectories separately.  We omit the standard deviation
if it is less than 1\%.}
\begin{center}
\begin{ruledtabular}
\begin{tabular}{llllll}
  &
  &
  \multicolumn{2}{c}{$\g=0,\ \b=9.6$} &
  \multicolumn{2}{c}{$\g=1/4,\ \b=5.6$}
\\
  \cline{3-4} \cline{5-6} & &
  accept & reject & accept & reject
\\ \hline
fat,high & max &  0.0783      & 0.0787      & 0.211       & 0.211      \\
         & avg &  0.0312      & 0.0312      & 0.0847      & 0.0848     \\
     & max/avg &  2.51        & 2.52        & 2.49        & 2.49       \\
\hline
fat,low  & max &  0.093(1)    & 0.091(2)    & 0.119(2)    & 0.112(4)   \\
         & avg &  0.0144      & 0.0144      & 0.0198      & 0.0198     \\
     & max/avg &  6.4         & 6.3(1)      & 6.0(1)      & 5.7(2)     \\
\hline
thin & max &  0.40(2)     & 0.80(8)     & 0.87(2)     & 0.95(7)    \\
     & avg &  0.0134      & 0.0134      & 0.0708      & 0.0708     \\
 & max/avg &  30(1)       & 60(6)       & 12.3(3)     & 13(1)      \\
\end{tabular}
\end{ruledtabular}
\end{center}
\end{table*}

We list the ensembles that we use for comparisons
in Table~\ref{tab:configs}.
This table also gives figures for the performance of the HMC algorithm
before and after adding the new term, to be discussed below.
In order to verify the similarity of the ensembles with and without
the new term, we present some results for particle spectra
and other physical quantities in Table~\ref{tab:spec}.

The crucial diagnostic information for two of our ensembles
is presented in Table~\ref{tab:impl}.
Let us begin with the data that pertain to the original action,
that is, $\g=0$.  These data, which are maximum and average impulse before
and after the chain rule, point to what has to be improved.
The first section of three rows, labeled as ``fat,high,'' provides information
on the fat-link impulse resulting from $S_{\textrm{high}}$.
Next, the ``fat,low'' section gives information on the fat-link
impulse from $S_{\textrm{low}}$.
Last, the ``thin'' section gives information on the thin-link impulse
resulting from the total fat-link impulse after the application of
the fat-to-thin chain rule.

In each section, the first row gives the
maximal impulse per link in each MD trajectory, averaged over trajectories,
separately for accepted and for rejected trajectories.
The next row similarly gives the average impulse.
The third row gives the ratio of mean maximal impulse to average impulse.

The main thing to notice about the fermions' fat-link impulses
of the $\g=0$ ensemble is that they are exactly the same
for accepted and for rejected trajectories.  In other words, there is
absolutely no correlation between the fermions' fat-link impulses
and the result of the Metropolis test of the trajectory.

By contrast, the thin-link impulses of the $\g=0$  ensemble
exhibit a clear distinction between accepted and rejected trajectories.
For accepted trajectories the max/avg ratio is about 30,
whereas for rejected trajectories it is twice as big.
Histograms of the maximal thin impulse can be found in the upper row
of Fig.~\ref{fig:fat}.  The difference in the mean value of the maximal
impulse between the left and right panels is clearly visible.
Also the shapes of the two distributions are quite different.

Our hypothesis is that when the bare coupling is strong enough,
dislocations in the dynamical gauge field become abundant.
Sometimes, such dislocations
will give rise to exceptionally small eigenvalues of the matrix $Q$
of Eq.~(\ref{Q}).  Through the fat-to-thin chain rule~(\ref{fatforce}),
the small eigenvalues of $Q$ generate spikes in the thin-link impulse,
which, in turn, results in a bigger probability for
failing the Metropolis test at the end of the trajectory.

It is obvious that a large impulse in the final, thin-link force
causes rejection of a trajectory.  What is new is our observation that
there is no large impulse in the initial, fat-link calculation.
Evidently the problem lies in the chain rule.
What contributes to the severity of this problem is that
even a single spike for a single link at a single update step
of the whole trajectory, if it is too big, has the potential of
producing such a violation of MD energy conservation that will
result in failing the Metropolis test.
The question is whether we can do something about it.

\section{Dislocation-suppressing action
  for \lowercase{n}HYP links\label{sec:nds}}
\dlbl{sec:nds}

In four dimensions, nHYP links $V_{x,\m}$ are constructed
from the dynamical gauge field $U_{x,\m}$ via three successive
smearing steps \cite{Hasenfratz:2001hp,Hasenfratz:2007rf}.
Each step consists of first constructing a weighted sum over staples,
which is then reunitarized.  Explicitly,
\begin{widetext}
\ulbl{smear}
\begin{subequations}
\begin{eqnarray}
  \O_{x,\r;\x} &=& (1-\a_3) U_{x,\r}
  + \frac{\a_3}{2} \left(
    U_{x,\x} U_{x+\hat\x,\r} U^\dagger_{x+\hat\r,\x}
    + U^\dagger_{x-\hat\x,\x} U_{x-\hat\x,\r} U_{x-\hat\x+\hat\r,\x}
  \right) \,,
\label{smeara}\\
  \rule{0ex}{3ex}
  \bV_{x,\r;\x} &=& \cp( \O_{x,\r;\x} ) \ ,
\NON
  \rule{0ex}{3ex}
  \bO_{x,\m;\n} &=& (1-\a_2) U_{x,\m}
  + \frac{\a_2}{4} \sum_{\stackrel{\scriptstyle \r \ne \m,\n}{\x \ne \m,\n,\r}}
    \left(
    \bV_{x,\r;\x} \bV_{x+\hat\r,\m;\x} \bV^\dagger_{x+\hat\m,\r;\x}
    + \bV^\dagger_{x-\hat\r,\r;\x} \bV_{x-\hat\r,\m;\x}
    \bV_{x-\hat\r+\hat\m,\r;\x}
  \right) \,,
  \hspace{5ex}
\label{smearb}\\
  \tV_{x,\m;\n} &=& \cp\Big(\bO_{x,\m;\n}\Big) \ ,
\NON
  \tO_{x,\m} &=& (1-\a_1) U_{x,\m}
  + \frac{\a_1}{6} \sum_{\n \ne \m} \left(
    \tV_{x,\n;\m} \tV_{x+\hat\n,\m;\n} \tV^\dagger_{x+\hat\m,\n;\m}
    + \tV^\dagger_{x-\hat\n,\n;\m} \tV_{x-\hat\n,\m;\n}
    \tV_{x-\hat\n+\hat\m,\n;\m}
  \right) \,,
\label{smearc}\\
  V_{x,\m} &=& \cp\Big(\tO_{x,\m}\Big) \ .
\nonumber
\end{eqnarray}
\end{subequations}
\end{widetext}
The reunitarization operator $\cp$ is defined in Eq.~(\ref{reu}).
Keeping track of this construction in reverse order, one can see
that the staple sum extends into a different direction at each smearing step.
The outcome is that a given fat link $V_{x,\m}$ depends
on a particular thin link $U_{y,\n}$ if and only if there exists a hypercube
to which both $V_{x,\m}$ and $U_{y,\n}$ belong.\footnote{%
  Like the original thin links, the nHYP links $V_{x,\m}$ reside in
  the fundamental representation.  In our work on higher-representation
  fermions we first construct the nHYP links $V_{x,\m}$, and then apply
  the appropriate group theoretic formulae to construct links in the
  desired representation from $V_{x,\m}$.
  This also adds a step to the chain rule in calculating the MD force.
}

We are now ready to introduce the dislocation-suppressing action
for nHYP smearing.  This is done by adding to the pure-gauge action $S_g$
a new term,
\ulbl{Sg}
\begin{equation}
  S_g = S_{\textrm{plaq}} + S_{\textrm{NDS}} \ ,
\label{Sg}
\end{equation}
where the new term is
\ulbl{dsa}
\begin{eqnarray}
  S_{\textrm{NDS}} &=& \frac{1}{2N_c} \sum_x \tr\!\left(
                  \g_1 \sum_{\m} \tQ_{x,\m}^{-1}
                + \g_2 \sum_{\m\ne\n} \bQ_{x,\m;\n}^{-1}\right.\nonumber\\
               && \left. + \g_3 \sum_{\r\ne\x} Q_{x,\r;\x}^{-1}
      \right) \ .
\label{dsa}
\end{eqnarray}
The motivation for introducing the nHYP Dislocation Suppressing action,
or NDS action for short, is clear.
The chain rule can produce spikes in the thin-link force
associated with small eigenvalues
of $Q_{x,\r;\x}$, $\bQ_{x,\m;\n}$ or $\tQ_{x,\m}$.  The NDS action
is designed to suppress them, by creating a repulsive potential
that is proportional to the sum of inverse eigenvalues of the $Q$ matrices.

If we were to add the NDS action $S_{\textrm{NDS}}$ to the usual plaquette action
while holding $\b$ fixed, we would be pushed back into weaker coupling,
and smaller lattice spacing.  From the
weak-coupling expansion $U_{x,\m} = \exp(iaA_{x\m})$ we obtain
the bare coupling as
\ulbl{gbare}
\begin{equation}
  \frac{1}{g_0^2} = \frac{\b}{2N_c} +
  \frac{1}{N_c} \left( \frac{\g_1 \a_1}{3} + \g_2 \a_2 + \g_3 \a_3
  \right) \ .
\label{gbare}
\end{equation}
The crucial question, which can only be addressed by performing numerical tests,
is whether the NDS action can improve the performance of the HMC algorithm
under the same physical conditions.  This question will be studied in
the next section.

In concluding this section we note that $S_{\textrm{NDS}}$ is easily implemented
in the existing code.  Using the generic notation of Sec.~\ref{sec:chain},
first, $Q^{-1/2}$ is needed for the construction of the nHYP links,
so one obtains $Q^{-1}=Q^{-1/2} Q^{-1/2}$ with basically no extra cost.
Also, for the calculation of the force, we have
\ulbl{drvQ}
\begin{equation}
  \frac{\partial}{\partial U}\, \tr Q^{-1}
  = 2\, \tr\! \left(Q^{-1/2}\, \frac{\partial Q^{-1/2}}{\partial U} \right) \ .
\label{drvQ}
\end{equation}
Once again, as can be seen from Eq.~(\ref{fatforce}), both $Q^{-1/2}$ and
$\partial Q^{-1/2}/\partial U$ have already been calculated,
and so it is trivial to obtain their product.

\section{Improvement of molecular dynamics update\label{sec:MD}}
\dlbl{sec:MD}

In our numerical work we use the following values
for the smearing parameters \cite{Hasenfratz:2001hp,Hasenfratz:2007rf}
\ulbl{alpha}
\begin{equation}
  (\a_1,\a_1,\a_1)=(0.75, 0.6, 0.3) \ .
\label{alpha}
\end{equation}
Also, we have limited our numerical tests of $S_{\textrm{NDS}}$ to the case
\ulbl{gamma}
\begin{equation}
  \g_1 = \g_2 = \g_3 = \g \ ,
\label{gamma}
\end{equation}
where the values of $\g$ are shown
in the first column of Table~\ref{tab:configs}.  We have two ensembles
without the NDS action, a $\b=9.65$ ensemble with a slightly weaker
bare coupling and a $\b=9.6$ ensemble with a slightly stronger
bare coupling; one ensemble with $\g=1/8$ and $\b=7.8$, and
one with $\g=1/4$ and $\b=5.6$.
In trying to achieve the same physical conditions one can never do
a perfect job at non-zero lattice spacing.
Still, the physical properties listed in Table~\ref{tab:spec}
show that all four ensembles exhibit reasonably similar physics.\footnote{%
  In the weak-coupling regime, it follows from Eqs.~(\ref{gbare})
  and~(\ref{alpha}) that $\b$ must be shifted by $-2.3\g$ in order
  to keep the bare coupling $g_0$ fixed.
  As can be seen from Tables~\ref{tab:configs} and~\ref{tab:spec},
  the actual $(\g,\b)$ pairs that produce roughly equal physics
  involve a much bigger shift in $\b$, showing that we are very
  far from the range of applicability of Eq.~(\ref{gbare}).
}

\subsection{The thin-link force and acceptance\label{sec:acc}}
\dlbl{sec:acc}

The first piece of evidence that the NDS action actually works is the
performance figures shown in Table~\ref{tab:configs}.
While maintaining the good acceptance rate intact,
we have been able to reduce $n_1$ from 16 to 12, and $n_2$ from 2 to 1,
which is a saving of more than 50\% in the number
of fermion inversions per trajectory.

In order to understand the origin of this improvement
we look at the impulses of the ensemble with ($\g$, $\b$)=(0.25, 5.6)
and compare them to those of the (0, 9.6) ensemble,
as shown in Table~\ref{tab:impl}.  First, like the (0, 9.6) ensemble,
also in the case of the (0.25, 5.6) ensemble the fat-link impulses
are the same for accepted and for rejected trajectories.
Once again, the Metropolis test is uncorrelated
with the fat-link impulses produced by the pseudofermions.

In comparing the actual values of the impulses between the two ensembles
we should keep in mind that the maximal and average impulses
reflect the different numbers of update steps chosen in the two case.
However, the different time increments cancel out in the max/avg ratios.
Indeed, these ratios turn out to be equal in all four cases:
they are basically the same for accepted and rejected trajectories,
as well as for the (0, 9.6) and (0.25, 5.6) ensembles.  This shows
that the fermion sector did not bias the
acceptance rate one way or another.

The main difference between the two ensembles is revealed in the
thin-link impulse, in the bottom section of Table~\ref{tab:impl}.
While in the case of the (0, 9.6) ensemble the max/avg ratios
were 30 and 60 for accepted and for rejected trajectories respectively,
in the case of the (0.25, 5.6) ensemble they are $\sim 12$ for both accepted and
rejected trajectories.  The NDS action has produced thin-link impulse ratios
that are, first, smaller, and second, uncorrelated with
the Metropolis test.  These features are also seen in the histograms
in the bottom panels of Fig.~\ref{fig:fat}.  Indeed, the two histograms
of the (0.25, 5.6) ensemble have essentially the same shape.

The conclusion is that, while maintaining roughly the same physical conditions,
the NDS action with $\g=1/4$ successfully removes virtually all of
the spikes of the thin-link impulse that resulted from the fat-to-thin
chain rule.  This is one of the main results of this paper.

A complementary observation is the following.
Unlike the maximal thin-link impulse, the average thin-link
impulse is the same for accepted and for rejected trajectories
in both the (0, 9.6) and (0.25, 5.6) ensembles.  However,
this average value is more than 5 times bigger in the case of
the (0.25, 5.6) ensemble than in the (0, 9.6) ensemble.
The removal of the high-end tail of the distribution of the thin-link impulse
by the NDS action has allowed us to increase the average impulse
(by decreasing the numbers of steps), without harming the acceptance rate.

\subsection{Integrator instability and ``safety trajectories''\label{sec:inst}}
\dlbl{sec:inst}

With $n_2=1$ in the $\g\ne0$ ensembles of Table~\ref{tab:configs},
an interesting practical question is how aggressively
can one reduce the number of steps of the outer level, $n_1$.
For example, how would the acceptance rate change if $n_1$ is further
decreased from 12 to 8?

We have carried out an exploratory study of this question
on ensembles with parameter values that are similar to (but not necessarily
identical with) those of the $\g\ne0$ ensembles of Table~\ref{tab:configs}.
Our main finding is that if we keep lowering $n_1$,
at some point we will run into a situation where the HMC update
experiences occasional, but long, sequences of rejections.
The obvious first thought would have been that the fat-to-thin chain-rule
spikes of the thin-link impulse are back.  However, an examination
of the pattern of impulses leads to a different picture.
First, the long sequences of rejections are typically characterized
by spikes in MD energy non-conservation as large as $\D S = O(100)$.
Second, an examination of the MD time histories reveals that the occurrence
of spikes of $\D S$ is virtually always correlated with
(much smaller) spikes of both the maximal and the average fat-link force
coming from $S_{\textrm{low}}$.

The conclusion is that we are looking at a familiar integrator instability.
The breakdown of the MD integration was nicely exemplified in
the case of a free harmonic oscillator
in Ref.~\cite{instabl}.
If $\o$ is the frequency of the oscillator, and $\d\t$ is the time
increment of the (leapfrog) update, the breakdown occurs when
the product $\o\,\d\t$ exceeds a critical value
that depends on the MD integration scheme.

In our simulations, the time increment $\d\t$ was held fixed.
Of course, since we are dealing with an interacting field theory,
many oscillators are present simultaneously,
and their frequencies are changing with the MD evolution.
In effect, there is therefore a maximal frequency
$\o_{max}$ that scales with $\l_{min}^{-2}$, where $\l_{min}^2$
is the smallest eigenvalue of $M^\dagger M$ (see Eq.~(\ref{Slow})).
We have looked at the low-lying spectrum of $M^\dagger M$
on some of our stuck streams and found that, indeed,
the rise in the fat-link force of $S_{\textrm{low}}$ is correlated with
the occurrence of an exceptionally small eigenvalue.  This, in turn,
gives rise to an exceptionally large value of the product $\o_{max}\,\d\t$,
and, ultimately, to the onset of an integrator instability
\cite{instabl}.

The alert reader would notice that the new problem
is itself a sign of success.
Indeed, it is the very smooth fat-link background,
provided by the NDS action, which allows for the Wilson matrix
to develop such small eigenvalues that are eventually capable of generating
integrator instabilities.

Various solutions to this problem exist in the literature.  First,
obviously, the simplest solution is to avoid reducing the number of steps
too much.  The high acceptance rates of the $\g\ne0$ ensembles reported
in Table~\ref{tab:configs} suggest that, with $n_1=12$, we did not
run into any integrator instabilities.  This is confirmed by an examination
of the histories of these runs.  The (0.25,5.6) ensemble shows no $\D S$
spikes at all.  Perhaps because of its smaller $\g$, the (0.125,7.8)
ensemble has a few
spikes, but none of them has generated a sequence of rejections.

One can do still better by adopting the strategy of Ref.~\cite{Jansen:1998mx}.
According to this strategy, one uses a relatively small number
of steps for most trajectories.  Every once in a while, a larger
number of steps is used for a ``safety trajectory.''  The idea is that,
in case the simulation has run into a sequence of rejections resulting from
an integrator breakdown, that sequence will terminate at the next
safety trajectory, where, thanks to its finer time increment, the
trajectory will (very likely) be accepted.

We have found that, as long as integrator instabilities are rare,
even a modest increase in $n_1$ is usually enough to eliminate all of them.
As an example, the already noted high acceptance rates
of the $\g\ne0$ ensembles of Table~\ref{tab:configs} suggest that
we might use $n_1=12$ only for the safety trajectories,
while using a smaller number of steps, perhaps $n_1=8$, for most trajectories.
The interval between two safety trajectories might be taken to be
5 or 10 trajectories.\footnote{%
  Reversibility of the MD update requires that the interval between
  two safety trajectories will be fixed beforehand.
}
The question of what is the optimal combination invites study,
but it is clear that the insertion of safety trajectories is a very cheap cure
for the instability problem.

\begin{table}[t]
\floatcaption{tab:ev}{Properties of the kernel operator \cite{ovkernel}.
The third column gives the average value of its 10 lowest eigenvalues.
The last column gives the number of matrix multiplications needed for
the construction of the overlap operator.
}
\begin{center}
\begin{ruledtabular}
\begin{tabular}{lldd}
$\g$ & $\b$ &
  \dttl{$\bar\l$}{4} &
  \dttl{$N_{\textrm{op}}\times 10^5$}{2}
\\ \hline
0     & 9.6  &  0.106(12) & 3.0 \\
0     & 9.65 &  0.163(11) & 1.7 \\
0.125 & 7.8  &  0.182(11) & 1.7 \\
0.25  & 5.6  &  0.322(15) & 1.1 \\
\end{tabular}
\end{ruledtabular}
\end{center}
\end{table}

\section{Improvement of valence chiral fermions\label{sec:ov}}
\dlbl{sec:ov}

Chiral fermions---domain-wall fermions and overlap fermions---are widely
used nowadays \cite{Kaplan:1992bt,DWF,Neuberger:1997fp}.
While domain-wall fermions are used both as dynamical
\cite{Arthur:2012opa,JLQCD} and as valence fermions,
overlap fermions are mostly used as valence fermions
(see, however, Ref.~\cite{dynoverlap}).

These chiral fermions are built from a kernel $K$,
which is supercritical Wilson-like (hermitian) operator.
Ideally, the kernel would have a spectral gap.
In reality, there is never a clean gap.
Instead the kernel operator has a mobility edge
that is at $O(1)$ in lattice units, with a localized spectrum below
the mobility edge and an extended spectrum above it \cite{localization}.
The near-zero spectrum of localized eigenmodes is always undesirable.
In the case of domain-wall fermions it is a dominant source
for the residual mass, which is a measure of the imperfection
of the chiral symmetry of the domain-wall system.
In the case of overlap fermions, such near-zero eigenmodes need
to be deflated during the construction of the overlap operator itself.
When more of them are present, this makes the numerical construction
more expensive and/or less accurate.

Since it is localized, a near-zero eigenmode of the kernel operator
often owes its existence to a dislocation
in the gauge field \cite{Berruto:2000fx}.
Now, the NDS action suppresses a certain family of dislocations, and so
it is interesting to study whether it has any effect on the behavior
of chiral fermions.  As we will see, we indeed find a clear improvement.

We have used nHYP links to construct the kernel operator introduced
in Ref.~\cite{ovkernel}, and studied its properties on our set of ensembles.
The third column of Table~\ref{tab:ev} gives the average value $\bar\l$
of the 10 lowest kernel eigenvalues $|\l_i|$, $i=1,\ldots,10$.
We see that $\bar\l$ grows with $\g$, and that, for $\g=1/4$, it is
significantly larger than in the other cases.  This shows that
the dislocation-suppressing effect of the NDS action also helps
in reducing the number of near-zero eigenvalues of the kernel operator.
This effect is also seen in Fig.~\ref{fig:ev}, which shows histograms
of the same 10 lowest kernel eigenvalues.
Moreover, the depletion of the near-zero
spectrum speeds up the numerical construction
of the (valence) overlap operator.
This can be seen in the last column of Table~\ref{tab:ev}, which shows
the average number of matrix multiplications by the kernel $K$
that is needed for the construction
process to converge to a given precision.  For valence
domain-wall fermions, we would correspondingly expect a reduction
of the residual mass at a fixed size of the fifth dimension.

\section{Discussion \label{sec:conc}}
\dlbl{sec:conc}

The chain rule relating a force with respect to nHYP links
to a force with respect to the dynamical links can give rise
to relatively rare, but large, spikes of the total impulse,
which, in turn, degrade the performance of the HMC algorithm.
In this paper we propose a new term in the gauge action
aimed to suppress such spikes.
We have shown that, for (approximately) fixed physical parameters,
the new term indeed improves the performance of the HMC algorithm.
As a side benefit, it improves the behavior of chiral valence fermions.

We have attributed the chain-rule spikes in the force
to dislocations in the gauge field---a somewhat vague term.
Indeed we have identified the dislocations, operationally,
by the very fact some of the $Q$ matrices (Eqs.~(\ref{Q}) and~(\ref{dsa}))
have exceptionally small eigenvalues.  This is analogous to what is
customary when dealing with domain-wall or overlap fermions,
where the presence of a dislocation is identified by the existence
of a near-zero (localized) eigenvalue in the spectrum of the kernel
operator.  What is common to both cases is that one is not interested
in the general roughness of the gauge field {\it per se},
but rather in concrete undesirable effects that this roughness
can produce.

Over the years, a large body of work has been devoted to improving
the performance of chiral fermions.  While it is beyond the scope of this paper
to review all this work, we would like to draw some useful lessons.\footnote{%
  The interested reader may consult
  Refs.~\cite{localization,Antonio:2008zz,DSDR} and references therein.
}
A common way to improve the behavior of chiral fermions
is to use so-called improved gauge actions.  Examples include
various variants of the Symanzik action, the Iwasaki action,
and the DBW2 action.
Each of these actions consists of a sum over a few Wilson loops.
The relative weight of each Wilson loop is fixed either by
finding an approximate solution of a truncated
renormalization group transformation, or by demanding the elimination
of the leading discretization effects perturbatively.
Generally speaking, these improved actions produce a smoother gauge field than
the simple plaquette action~(\ref{Splaq}).  The decreased
roughness of the gauge field typically gives rise to fewer near-zero modes
in the kernel's spectrum \cite{localization,Antonio:2008zz}.

One method designed to suppress the near-zero eigenvalues
of the chiral fermions' kernel $K$
is the so-called Dislocation-Suppressing Determinant Ratio (DSDR)
\cite{Arthur:2012opa,DSDR}.
The basic idea is that, if we were to add to the gauge action the term
\ulbl{dsdr}
\begin{equation}
  -\log\det(K^\dagger K) \ ,
\label{dsdr}
\end{equation}
this would produce a logarithmically divergent repulsive potential
that entirely suppresses any exact zero modes of the kernel operator.
In practice, using the term~(\ref{dsdr}) also has undesirable effects,
and so, instead, one replaces $K^\dagger K$ in the above expression
by a certain rational polynomial of $K^\dagger K$.

The NDS action~(\ref{dsa}) is analogous to DSDR in that it targets
those dislocations that are responsible for a specific undesirable feature.
Now, such dislocations represent local, lattice-size structures
in the dynamical gauge field that do not scale. Hence, there is no
particular reason to fix the weight of the dedicated, NDS term relative to
other terms in the gauge action, and it might be more natural
to hold fixed the absolute coefficient of the NDS term while varying
the coefficient of the plaquette action (or of any improved action,
if one is being used).  This is what is being done in effect in the
domain-wall simulations of the RBC and UKQCD collaborations:
a fixed DSDR term is used
while the coefficient of the Iwasaki gauge action is being varied.

Such technical details need not obscure the basic fact that all of the
various types of improvement usually play in concert, as was
found in the context of domain-wall and overlap fermions
\cite{Arthur:2012opa,DSDR}.
A new example is what we have found in this paper: The NDS action,
designed specifically to remove chain-rule spikes in the force for nHYP
smearing, also reduces the density of near-zero kernel
eigenvalues for chiral fermions.

An alternative way to avoid the chain-rule spikes of the force
is to reduce the values of the smearing parameters [see Eq.~(\ref{alpha})].
It was observed in Ref.~\cite{Cheng:2011ic} that the matrix $Q$ is
positive definite if all the smearing parameters are smaller
than 0.5.  The dislocations will still be there, however, and their effect
on chiral valence fermions will be undiminished.  Moreover, weakening the
smearing will destroy some of its benefit for approaching the continuum
limit.

We mentioned HISQ fermions \cite{HISQ,HISQ2},
which also incorporate the reunitarization step~(\ref{reu}).
Indeed the same correlation
between small eigenvalues of the matrix $Q$ and peaks of the MD force
has been reported in this case.\footnote{%
  See Appendix B  of Ref.~\cite{HISQ2}.
}
It is likely that building
a dislocation-suppressing action adapted for this type of smearing
would similarly improve the performance of HISQ simulations.

\begin{acknowledgments}
We thank Anna Hasenfratz for useful discussions.
This work was supported in part by the Israel Science Foundation
under grant no.~449/13, and by the U.~S. Department of Energy
under grant DE-FG02-04ER41290.
Our computer code is based on the publicly available package of the
MILC collaboration~\cite{MILC}.
The code for hypercubic smearing was adapted from a program written
by A.~Hasenfratz, R.~Hoffmann and S.~Schaefer~\cite{Hasenfratz:2007rf}.
\end{acknowledgments}


\clearpage
\pagebreak
\begin{figure*}
\begin{center}
\includegraphics[width=.47\textwidth,clip]{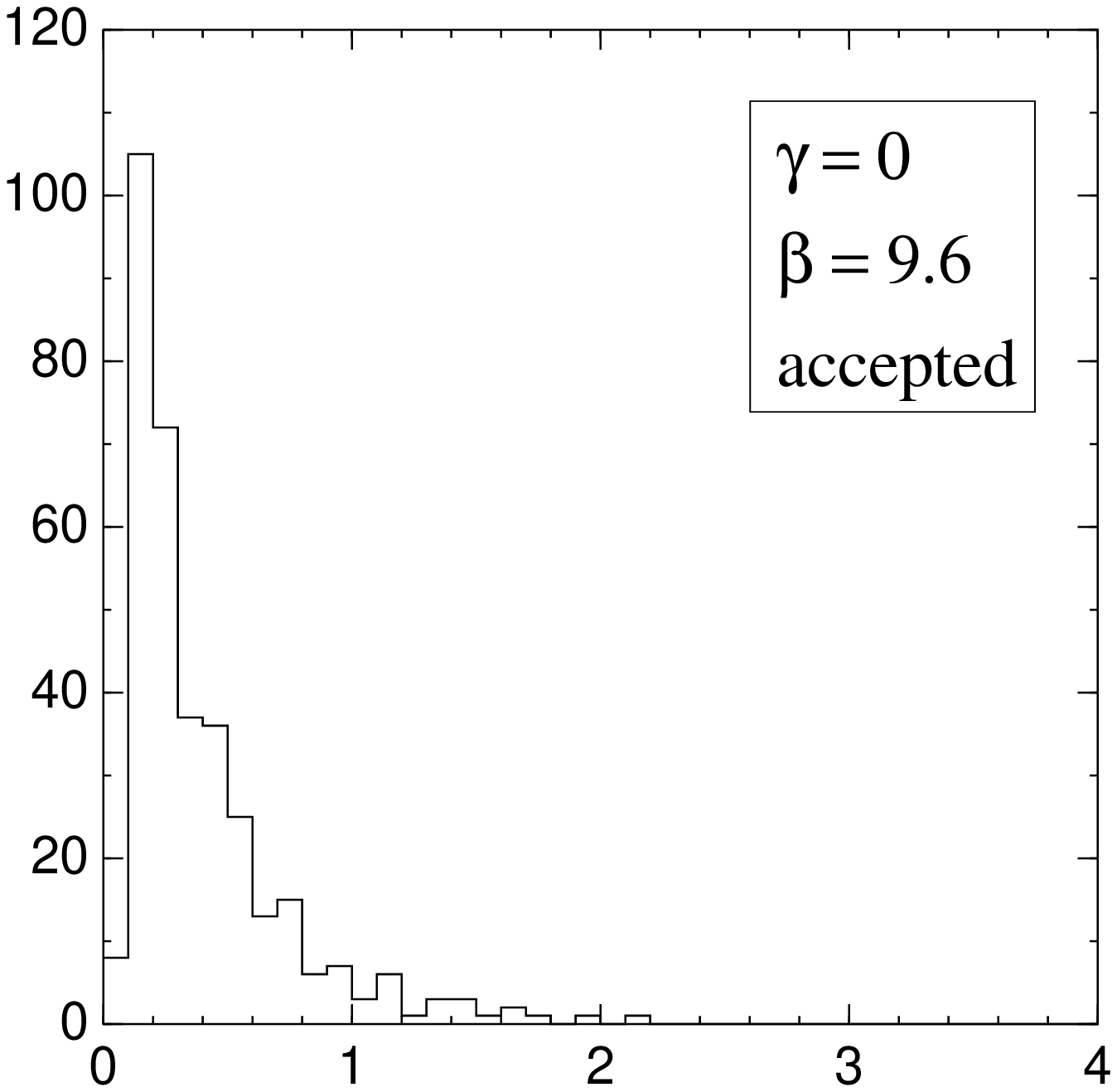}
\includegraphics[width=.47\textwidth,clip]{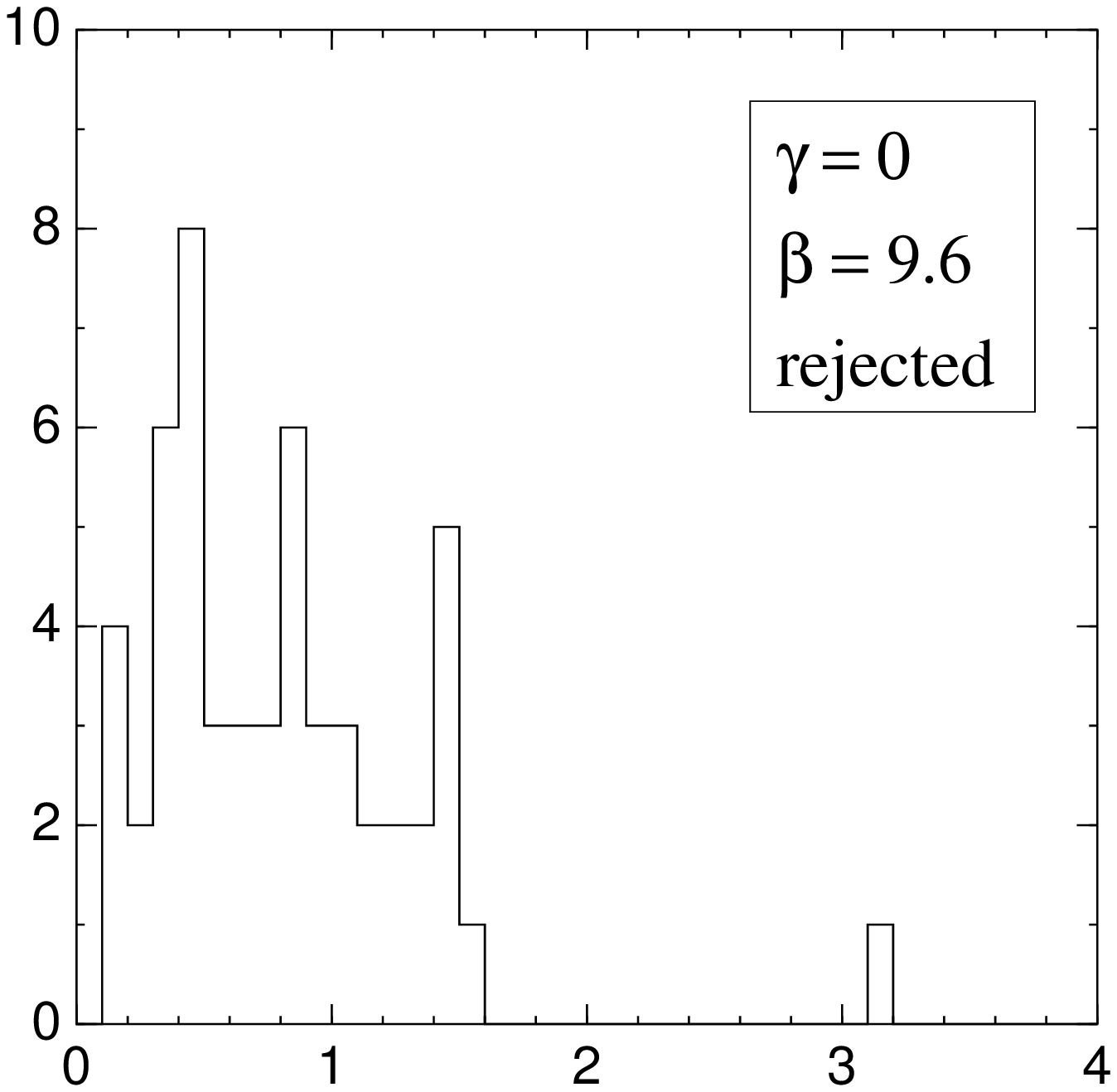}\\
\includegraphics[width=.47\textwidth,clip]{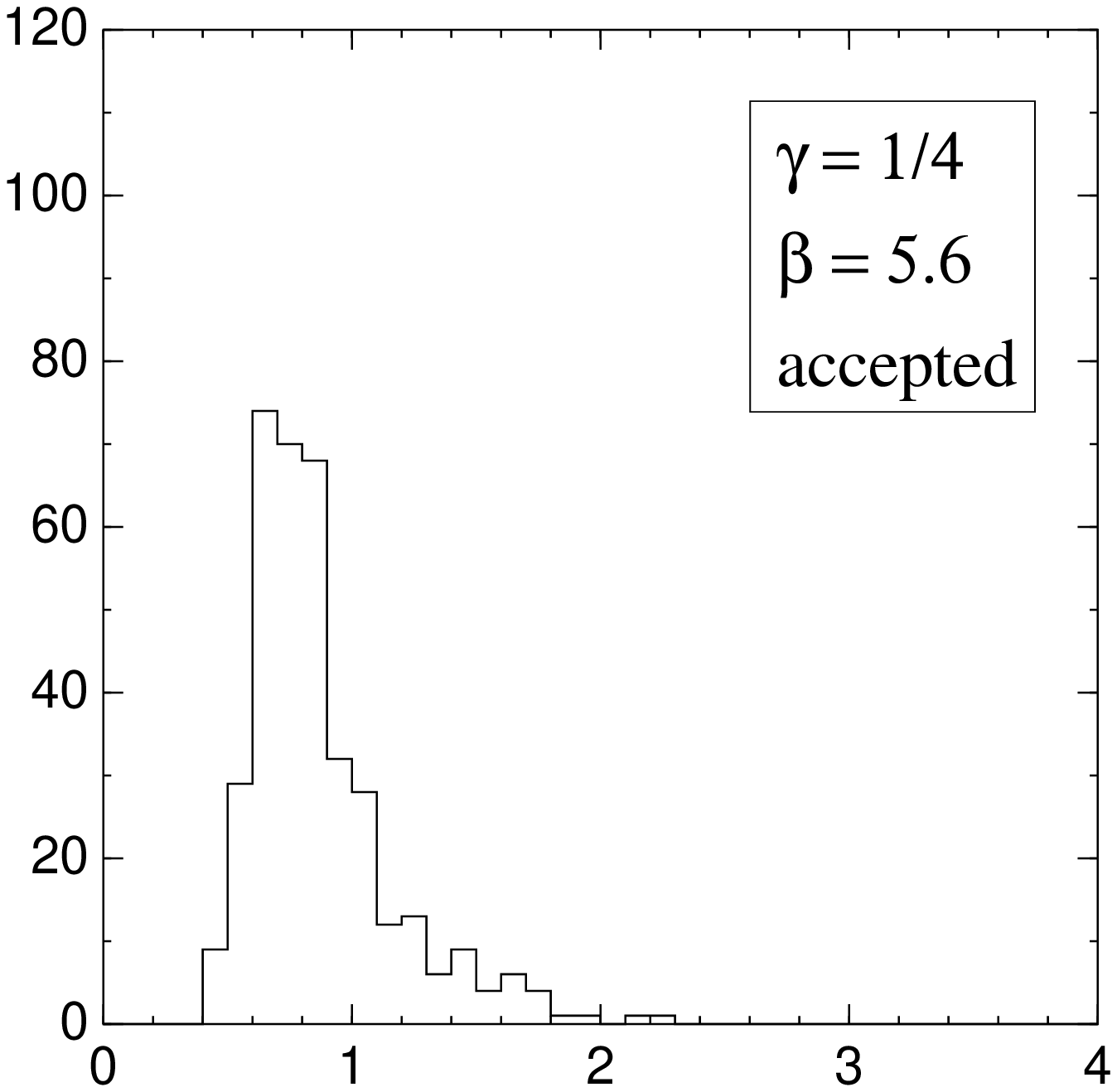}
\includegraphics[width=.47\textwidth,clip]{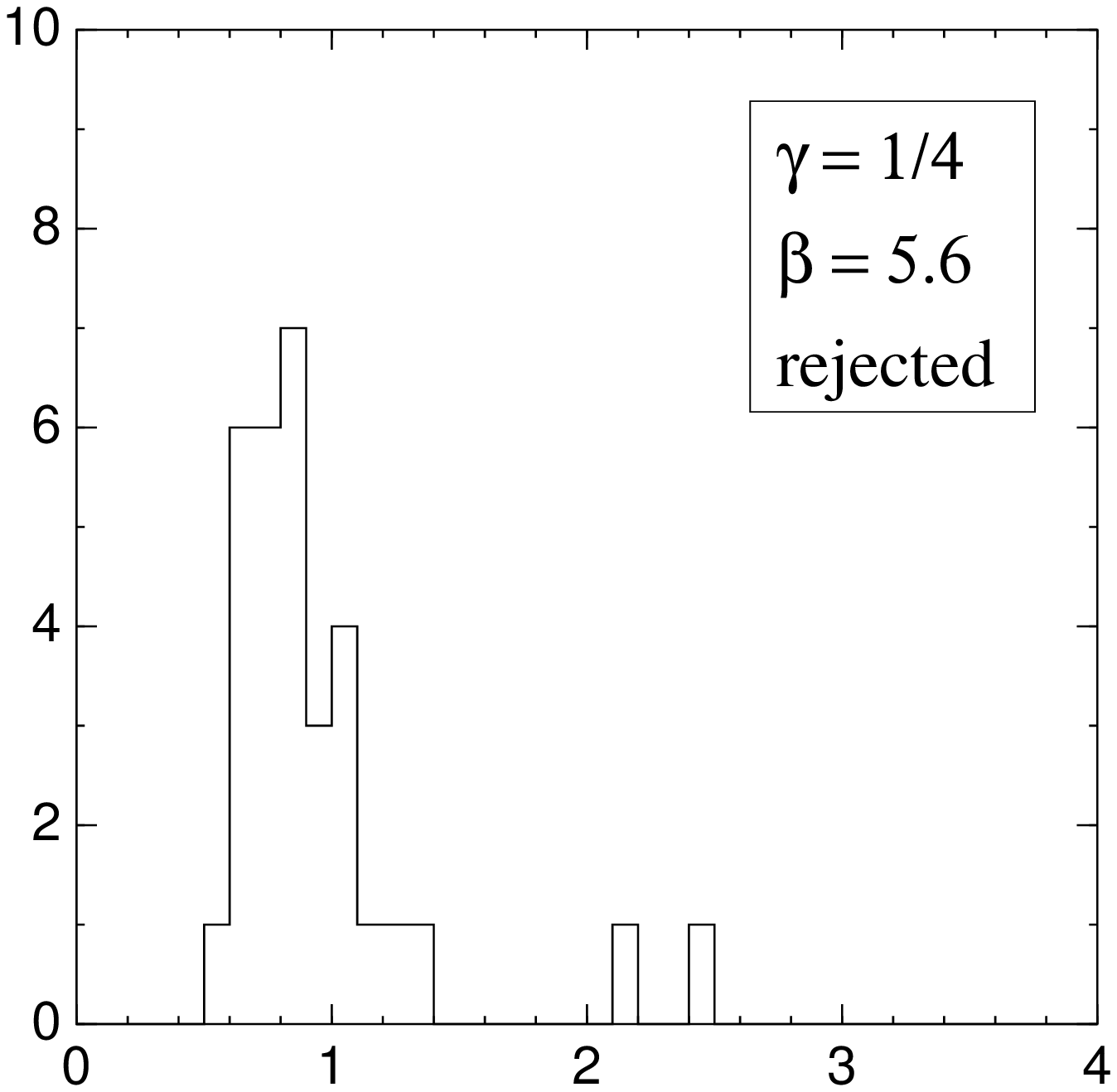}
\end{center}
\floatcaption{fig:fat}{Histograms of the maximal thin-link force
(corresponding to the ``thin'' section of Table~\ref{tab:impl}),
for 400 trajectories each.  Notice the different vertical scales
for accepted (left) and for rejected (right) trajectories.}
\end{figure*}

\begin{figure*}
\begin{center}
\includegraphics[width=.2\textwidth,clip]{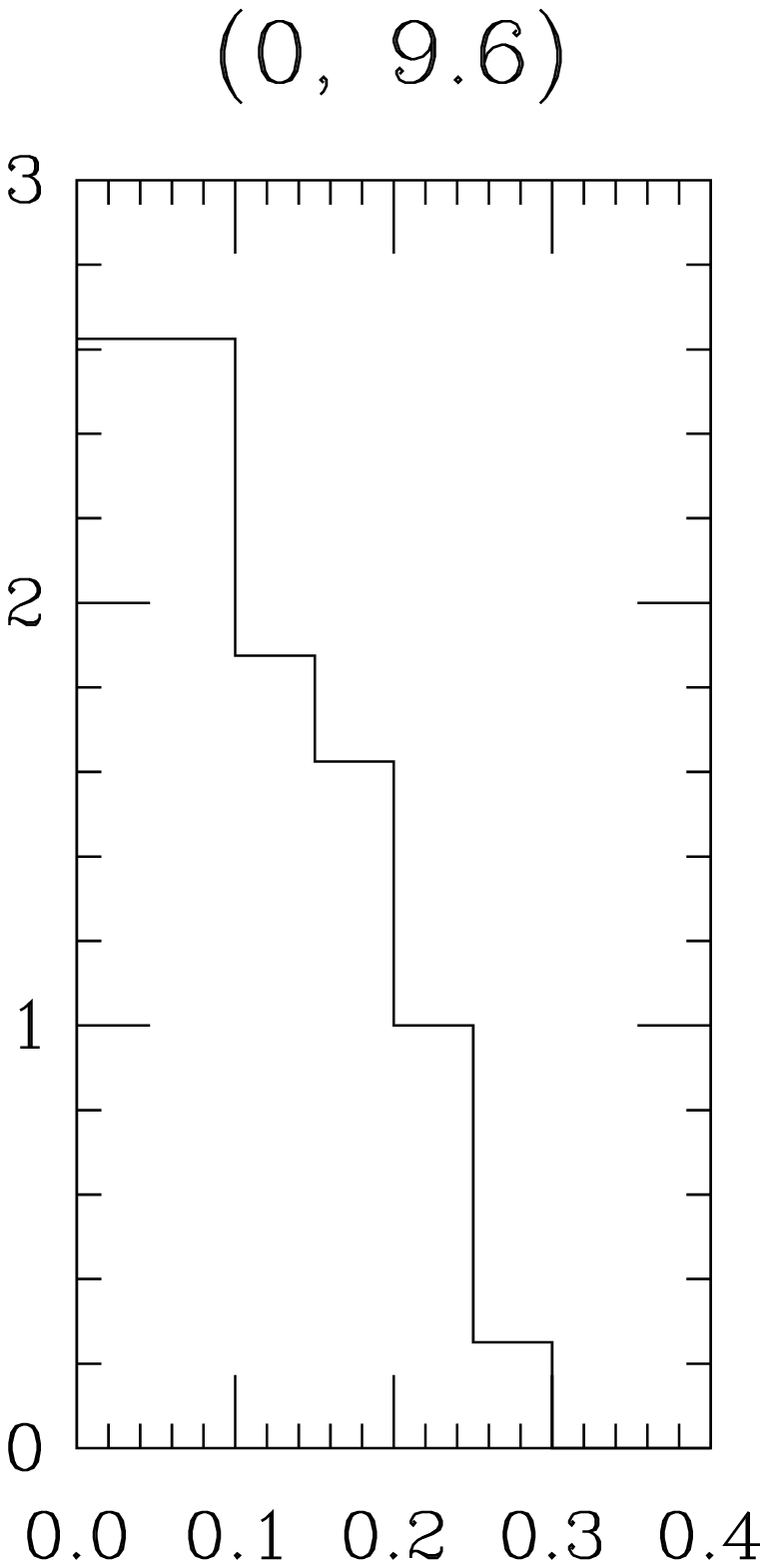}\hspace{-.5ex}
\includegraphics[width=.2\textwidth,clip]{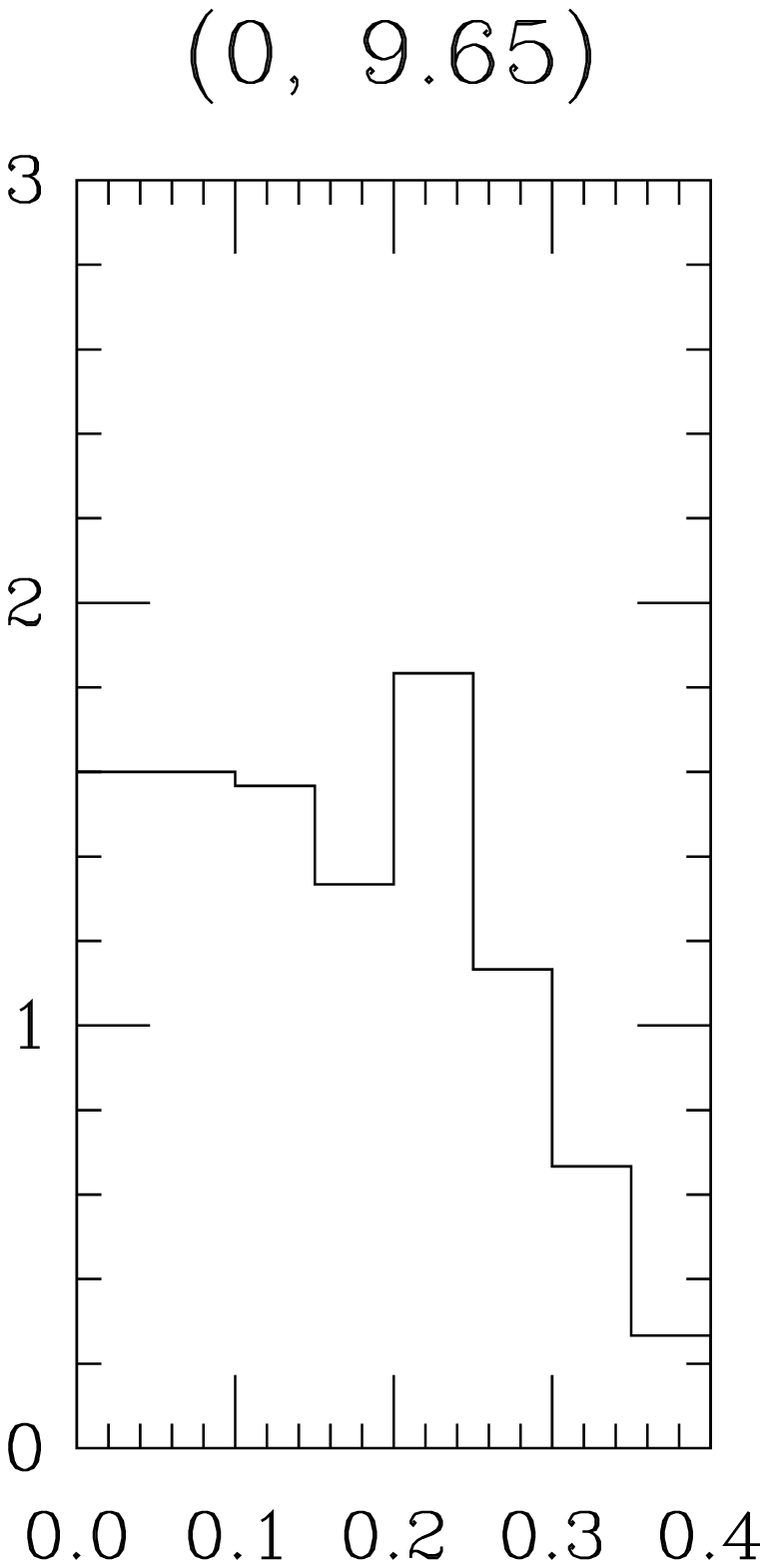}\hspace{-.5ex}
\includegraphics[width=.2\textwidth,clip]{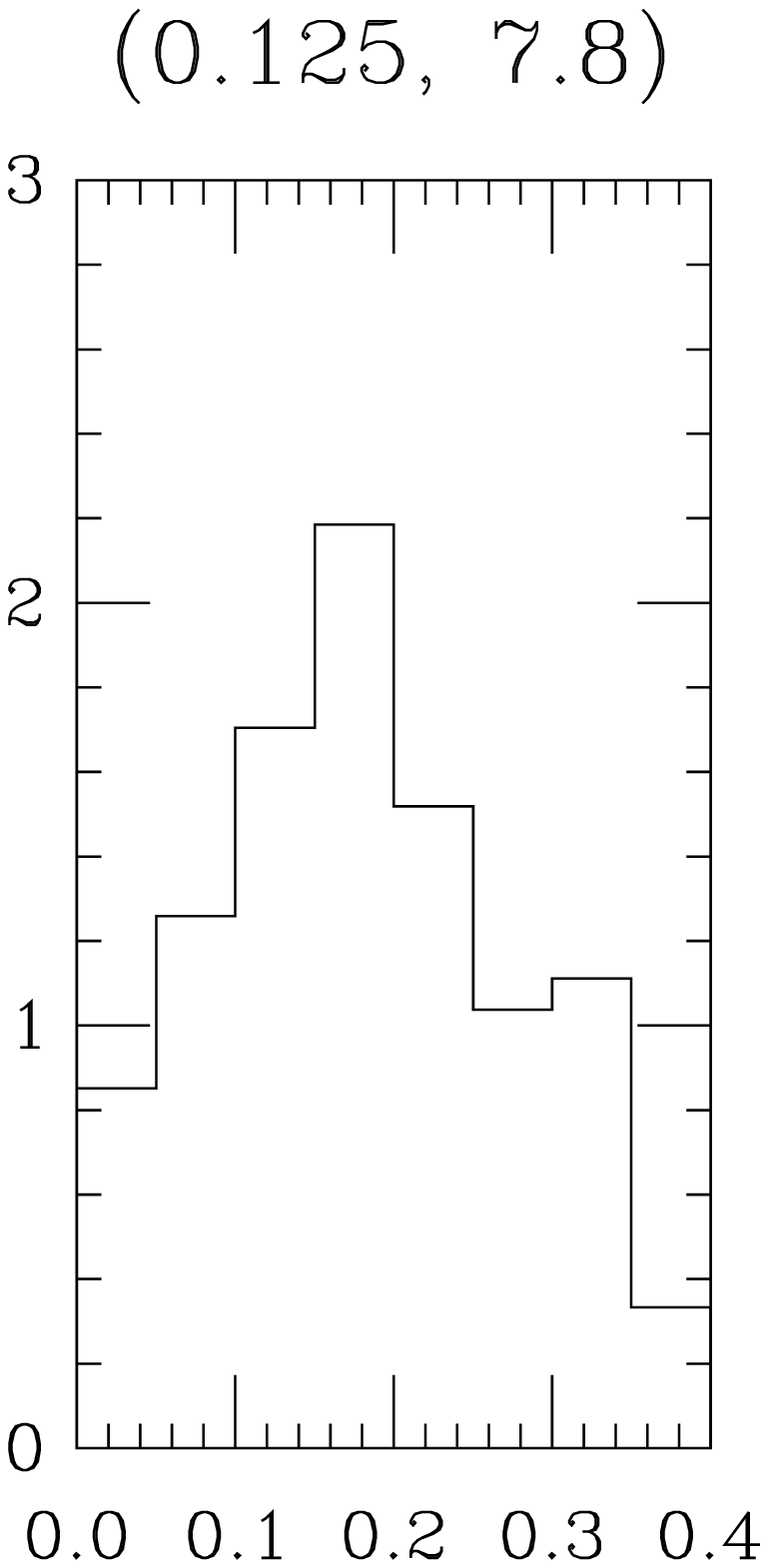}\hspace{-.5ex}
\includegraphics[width=.2\textwidth,clip]{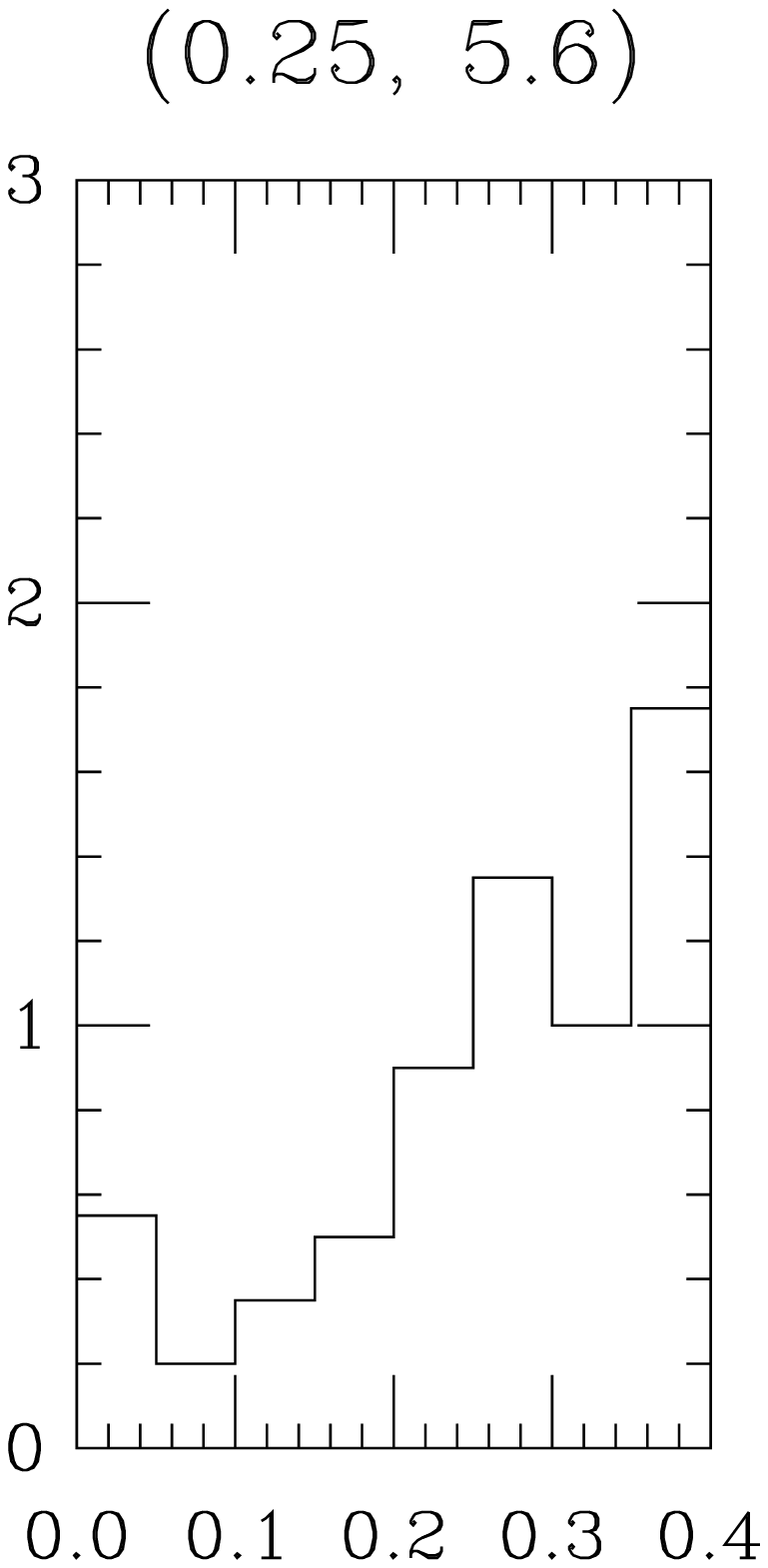}
\end{center}
\floatcaption{fig:ev}{Histograms of the lowest 10 eigenvalues of the
kernel operator.  The vertical axis is the average number of
eigenvalues per bin, where the bin size is 0.05.
For the ($\g$, $\b$)=(0, 9.6) ensemble,
the lowest 10 eigenvalues of all the configurations were within the shown
interval, whereas for the other cases, some of these eigenvalues
fall outside of this interval.  Notice the depletion of near-zero
eigenvalues as $\g$ is increased.}
\end{figure*}

\end{document}